\documentstyle[psfig,aps,prb,multicol]{revtex}

\def\mb#1{\mbox{\boldmath $#1$}}
\def\mbscr#1{\mbox{\scriptsize\boldmath $#1$}}
\def\scr#1{\mbox{\tiny #1}}

\title{Coherent Umklapp  Scattering of Light from 
Disordered Photonic Crystals}

\author{A. Yu. Sivachenko, M. E. Raikh, Z. V. Vardeny}
\address{Department of Physics, University of Utah, Salt Lake City 
UT 84112}
\date{\today}
\begin{document}

\maketitle

\begin{abstract}
A theoretical study of the coherent light scattering from disordered
photonic crystal is presented.
In addition to the conventional enhancement of the reflected 
light intensity
into the backscattering direction, the so called coherent
backscattering (CBS), the periodic modulation of 
the dielectric function in photonic crystals
gives rise to a qualitatively new effect: enhancement of the reflected
light intensity in directions different from the backscattering
direction. 
These additional coherent scattering processes, dubbed here
{\em umklapp scattering} (CUS), result in peaks, which are 
most pronounced when the incident light beam enters the sample at
an angle close to the
the Bragg angle. 
Assuming that the dielectric function modulation is weak, 
we study the shape of the CUS peaks
for different relative lengths
of the modulation-induced Bragg attenuation compared to 
disorder-induced mean
free path.  We show that when the Bragg 
length increases, then the CBS peak assumes its conventional shape, 
whereas the CUS peak rapidly diminishes in amplitude. 
We also study the suppression of the
CUS peak upon the departure of the incident beam
from Bragg resonance: we found that the diminishing of the CUS intensity 
is accompanied by substantial broadening. 
In addition, the peak becomes asymmetric.
\end{abstract}

\pacs{PACS numbers:71.55.Jv,42.70.Qs,42.25.Dd}

\begin{multicols}{2}

\section{Introduction}

Since the first experimental 
observations\cite{lagexp1,wolf,akkexp,andrejco,kaveh,lagexp2,etemad,akkjphys},
the phenomenon of coherent backscattering (CBS) of light
from disordered media has been the subject of intense 
theoretical and experimental studies\cite{sheng} 
(see also Ref. \onlinecite{review} for the most recent review). 
The underlying
mechanism for the CBS was identified as interference of clockwise
and counterclockwise scattering paths. This was understood already
in the early works by analogy to weak localization of electrons.
It has been also pointed out\cite{John} that this picture 
{\em fully} captures the 
physics of the coherent scattering only if 
there are no forbidden directions for the propagation of light
in the absence of disorder.
These forbidden directions emerge in systems with  periodic spatial
modulation of the dielectric function, or in other words, in photonic
crystals with incomplete gaps\cite{Joannopoulos}. 
In the presence of periodicity, the enhanced  scattering of light 
may occur not only in the backscattering direction, but in other  
directions as well. Roughly speaking, the additional peaks in  the  
scattering intensity can be regarded as
periodicity-induced diffraction satellites of the CBS peak. 
Their origin is illustrated in Fig.~1. In the presence of the periodic
modulation of the dielectric function, the light wave vector is determined
only up to the vector $\mb{\sigma}$ of the reciprocal lattice. Hence, 
upon entering the medium, light
with wave vector $\mb{k}$ acquires a satellite component with $y$ 
projection of the wave vector, $k_y=k\sin\theta-\sigma$. Importantly,
the same argument also applies for the
coherently backscattered light with wave vector
$-\mb{k}$, when it propagates inside the medium on the way out, namely
it also acquires a component
with $y$ projection $k_y=-k\sin\theta+\sigma$ (Fig. 1). 
This component 
gives rise to a satellite of the CBS peak in the direction $\theta'$ 
where $\sin\theta'=(\sigma-k\sin\theta)/k$. 
We dub this 
\parbox[t]{8.65cm}{
\noindent
\psfig{figure=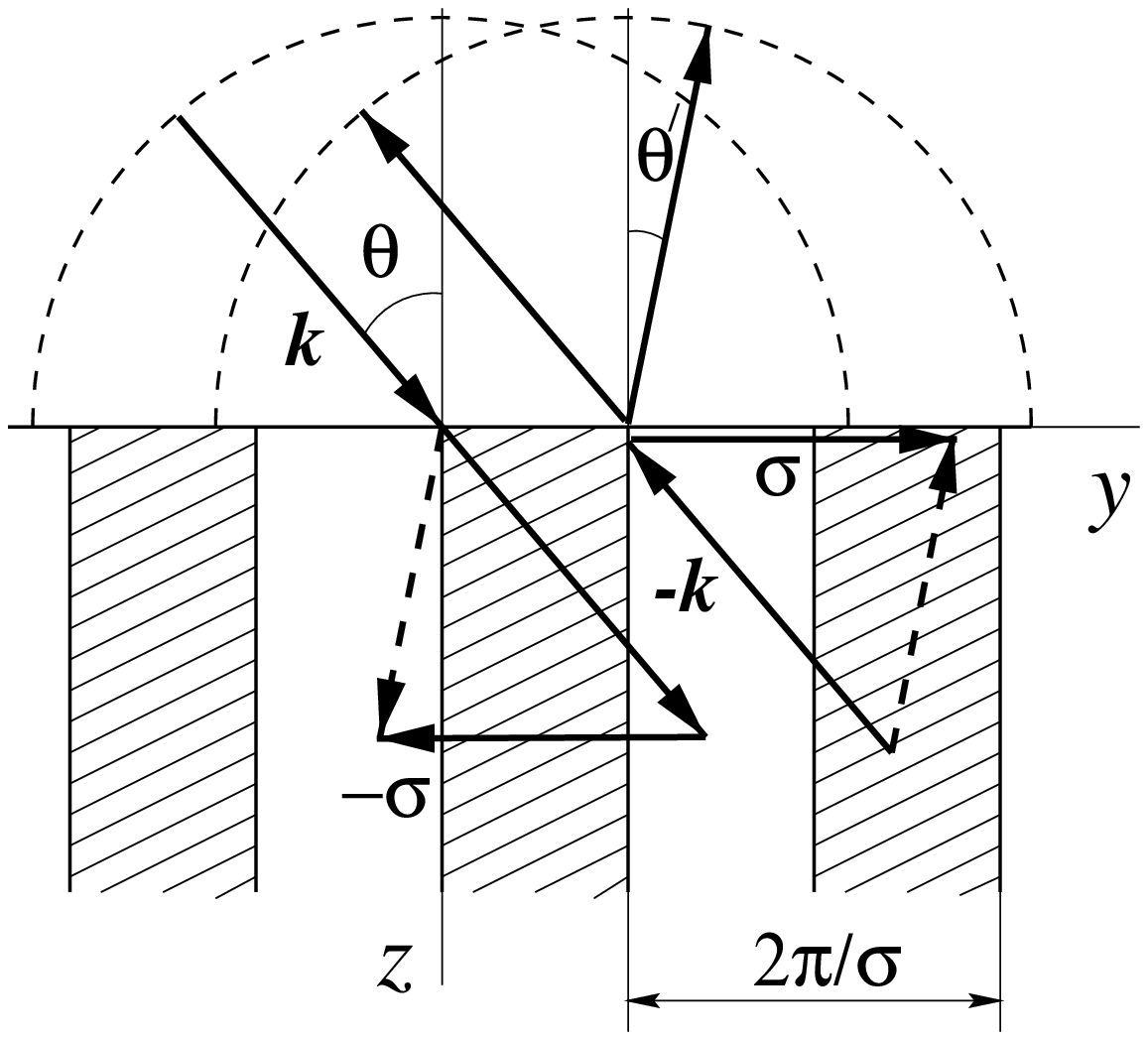,width=8.65cm}

\vskip2mm

\noindent
\small {\bf Fig. 1.} Schematic plot of light diffraction in a photonic 
crystal. The diffraction satellites of the incident (at angle $\theta$)
and outgoing plane waves are shown with dashed lines. 
The diffraction satellite of the backscattered  wave 
({-\mbox{$k$}}) 
results in outgoing wave emerging at an angle 
$\theta'$.

\vskip3mm
}
peak in the reflected light intensity as coherent {\em umklapp}
scattering (CUS). We note that the above picture is only illustrative. 
In reality, in addition to the process shown in Fig.~1, a variety of 
diffraction processes contribute to the formation of the CUS peaks. 
In general, the total number of the CUS peaks is 
determined by the number of reciprocal lattice vectors for which 
$\vert -\mb{k}+\mb{\sigma}\vert<k$.

It follows from the above qualitative picture that the magnitude 
of the CUS peak is governed by the ratio 
of the disorder-induced mean free path, $l$ (the elementary step of 
light diffusion) and the characteristic length, $\xi$, of formation 
of the diffraction component. 
This formation occurs most efficiently for the 
incidence angles 
$\theta=\theta_{\scr{B}}$, 
corresponding to the Bragg condition $k_y=\sigma/2$, {\em i.e.}
$\sin\theta_{\scr{B}}=\sigma/(2k)$. 
In this case $\xi$ coincides with the Bragg decay length, 
$L_{\scr{B}}$, which is the decay length of evanescent 
wave with frequency 
in the middle of the photonic stop band. Away 
from Bragg resonance, the length $\xi$ increases
resulting in suppression of the CUS peak. We note that the limiting case
 $L_{\scr{B}}\rightarrow 0$ and $\theta=\theta_{\scr{B}}$ was considered
in Ref.\onlinecite{Gorodnichev}.  In this 
case the CBS and CUS peaks are simply
the mirror images of each other. 
In the case of photonic 
crystal with incomplete gap, $L_{\scr{B}}$ is large 
($kL_{\scr{B}}\gg 1$). The 
question of interest is then: how do the magnitude and shape of the 
CUS peak depend 
on $L_{\scr{B}}/l$ and on the deviation of the incident beam from the Bragg 
angle. This question
is studied in the present paper. We generalize the approach of 
Refs.~\onlinecite{akkexp} and~\onlinecite{Akkermans} to the case of 
periodically modulated dielectric medium
and derive analytical expressions for the CBS and CUS peaks. 
We trace the evolution of the amplitude and width of the CUS peak
with changing the parameter $L_{\scr{B}}/l$ and the incidence angle,
$\theta$. We found that the most pronounced variation occurs within the domain 
$L_{\scr{B}}\lesssim l$ and 
$(\theta-\theta_{\scr{B}})\lesssim (kL_{\scr{B}})^{-1}$.

The paper is organized as follows. In Section II we review
the conventional derivation of the CBS and introduce the modifications 
necessary to take into account the Bragg reflection in periodic
structure. In Sec. III we derive analytical expressions for the CUS
and CBS albedo from disordered photonic crystal and analyze limiting
cases. In concluding remarks, Sec. IV, we outline generalization
of our theory to the case of arbitrary direction of the modulation
wave vector. 

\section{Coherent albedo}

\subsection{Coherent backscattering from disordered medium}

\vskip-1mm

In order to introduce the notations used throughout this paper, 
we first review
the conventional derivation of the CBS following 
Ref.~\onlinecite{Akkermans}. Neglecting the polarization 
effects\cite{stephen},
the intensity $I(\mb{R})$ reflected from the medium when illuminated with
incident flux $F_0$ and which is observed at
point $\mb{R}$, is given by the well known expression\cite{Akkermans},
\begin{eqnarray}\label{albedo}\nonumber
I(\mb{R})&=&F_0\int G(\mb{R},\mb{r}_1') G^*(\mb{R},\mb{r}_2')
U(\mb{r}_1^{},\mb{r}_1',\mb{r}_2^{},\mb{r}_2') \\
&\times&\Psi_{\scr{inc}}(\mb{r}_1)
\Psi^*_{\scr{inc}}(\mb{r}_2)\;d^3r_1d^3r_2d^3r_1'd^3r_2',
\end{eqnarray}
where $G(\mb{R},\mb{r})$ is the mean propagator from $\mb{r}$ (inside
the medium) to the observation point $\mb{R}$, 
$\Psi_{\scr{inc}}(\mb{r})$
is the  
normalized mean incident field at point $\mb{r}$ inside the 
medium,
and $U(\mb{r}_1^{},\mb{r}_1',\mb{r}_2^{},\mb{r}_2')$ is the sum of all 
scattering diagrams with ends stripped\cite{Akkermans}. 

There are two leading contributions to 
$U(\mb{r}_1^{},\mb{r}_1',\mb{r}_2^{},\mb{r}_2')$. The first one 
comes from the ladder
diagrams and describes the background incoherent scattering. 
The second contribution, which is
responsible for the coherent enhancement
of scattered intensity, represents the sum of 
maximally crossed diagrams
\begin{equation} \label{ui}
U_i
=\frac{4\pi c}{l^2}
P(\mb{r}_1^{},\mb{r}_1')
\delta(\mb{r}_1^{}-\mb{r}_2')\delta(\mb{r}_1'-\mb{r}_2^{}),
\end{equation}
where $P(\mb{r}_1^{},\mb{r}_1')$ is the stationary probability
distribution to travel diffusively from
$\mb{r}$ to $\mb{r}_1'$ inside the scattering medium. For the disordered
medium with a boundary, 
this probability distribution is given by
\begin{equation}\label{pdiff}
P(\mb{r}_1^{},\mb{r}_1')=\frac{1}{4\pi D|\mb{r}_1^{}-\mb{r}_1'|}-
\frac{1}{4\pi D|\mb{r}_1^{}-\mb{r}_1^*|},
\end{equation}
where $D=lc/3$ is the light diffusion constant and $c$ is the speed of light. 
The first term in
Eq.~(\ref{pdiff}) is a conventional propagator in
the bulk medium. The second term 
ensures the boundary condition $P(z_0^{},\mb{r}_1')=
P(\mb{r}_1,z_0)=0$ that is imposed to describe the 
diffusion inside a semi-infinite medium.
The point $\mb{r}^*_1$ is the mirror image of the point
$\mb{r}_1'$ with respect to the trapping plane
located at $z_0\approx-0.7 l$ (see Ref.~\onlinecite{review} and
references therein).

The mean incident field in the case of translationally
invariant (on average) system is given by
\begin{equation} \label{psi}
\Psi_{\scr{inc}}(\mb{r}_1)=\exp\left(-\frac{z_1}{l\cos\theta}
+i\mb{k}\mb{r}_1\right),
\end{equation}
where $\mb{k}$ is the wave vector of the incident light 
($|\mb{k}|=k=\omega/c$) and $\theta$ is the angle of
incidence (Fig.~1). The first
term in the exponent describes the 
decay of the incident
mean field amplitude due to scattering.

For the observation point $\mb{R}$ in the far field region, 
the asymptotic
expansion of Green's function $G(\mb{R},\mb{r}_1)$ is
\begin{equation}\label{g}
G(\mb{R},\mb{r}_1)\approx \frac{e^{ikR}}{4\pi R}\;
\exp\left(-\frac{z_1}{l\cos\theta'}-i\mb{k}'\mb{r}_1\right),
\end{equation}
where $\mb{k}'$ is the wave vector in the direction of observation, 
$|\mb{k}'|=\omega/c$, and $\theta'$ is the angle between
$\mb{k}$ and $z$ axis.

Substitution of Eqs.~(\ref{ui})--(\ref{g}) into Eq.~(\ref{albedo}) 
results in the following expression for the CBS albedo, $\alpha$,
defined as the scattered intensity divided by the incident flux and 
the sample area $S$, 
\end{multicols}
\vskip-5.5mm
\hbox to8.9cm{\hrulefill\vrule height2mm}
\vskip-3mm
\begin{eqnarray} \nonumber 
\alpha(\mb{k},\mb{k}')&=& \frac{3}{(4\pi)^2Sl^3}
\int d^2\mb{\rho}dz_1 dz_1'\;  \exp\left[i\mb{q}\mb{\rho}+
i\kappa(z_1-z_1')-b\frac{(z_1+z_1')}{l}\right] \\ \label{aint}
&\times&\left(\frac{1}{\sqrt{\rho^2+(z_1-z_1')^2}}-
\frac{1}{\sqrt{\rho^2+
(z_1+z_1'+2z_0)^2}}
\right),
\end{eqnarray}
\begin{multicols}{2}
where $\mb{\rho}$ is the component of $(\mb{r}_1-\mb{r}_1')$ 
parallel to the medium boundary, $\kappa=(\mb{k}+\mb{k}')_z$, 
\mbox{$\mb{q}=\{(\mb{k}+\mb{k}')_x,(\mb{k}+\mb{k}')_y\}$}, and 
$b=(1/\cos\theta+1/\cos\theta')$.
Due to the presence of the fast oscillating exponent, 
$\exp(i\mb{q}\mb{\rho})$, the integral in Eq.~(\ref{aint}) is 
non-zero only within a narrow interval $|\theta'-\theta|\sim(kl)^{-1}$
around the backscattering direction, {\em i.e.} $\mb{q}=0$. In 
Ref.~\onlinecite{Akkermans} this integral
was evaluated analytically for small angles $\theta$, $\theta'$.
In fact, a general expression that is valid for arbitrary $\theta$, $\theta'$ 
can be obtained,
\begin{eqnarray}\label{fdef} \nonumber
\alpha(\mb{k},\mb{k}')&=& f(\kappa,\mb{q},b)=\frac{3}{8\pi S}\;
\frac{1}{(\kappa l)^2+
(b+ql)^2} \\
&\times&
\left[\frac{1}{b}+\frac{1-\exp(-2qz_0)}
{ql}\right].
\end{eqnarray}
In the next subsection we trace how the albedo~(\ref{fdef}) is 
modified in the presence of a weak periodic modulation of the
dielectric function inside the scattering medium.

\subsection{Coherent scattering in the presence of a photonic crystal}

\subsubsection{Modification of the wave amplitudes}

Our main observation is that incomplete photonic gap affects
light diffusion only weakly\cite{Raikh}. The probability
that a certain step of the diffusive motion occurs in the
forbidden direction can be estimated as 
$(kL_{\scr{B}})^{-1}\ll 1$. On the other hand, {\em before the
first} and {\em after the last} scattering events the light
propagates along fixed directions $\theta$ and $\theta'$.
If either $\theta$ or $\theta'$ is close to the forbidden
direction, then light propagation will be strongly affected
resulting in a significant change in the albedo.
This observation suggests that in order to calculate 
the coherent albedo from a photonic crystal with incomplete
gap, it is sufficient to modify only the incident field 
amplitude in Eq.~(\ref{psi})
and the Green function of emerging light in the Eq.~(\ref{g}), without 
changing the propagator in Eq.~(\ref{ui}).

To model the incomplete band gap, which is narrow compared
to the Bragg frequency one can keep only a single harmonics
in the spatial modulation of the dielectric function,
\begin{equation}\label{epsilon}
\delta\varepsilon(y)=2\delta\varepsilon\;\cos(\sigma y).
\end{equation}
Here we assumed for simplicity that the direction of modulation
is parallel to the boundary as shown in Fig.~1. 
Generalization of the results
to an arbitrary angle between the boundary and the modulation
wave vector is outlined in the concluding remarks.

As was pointed out above, the Bragg resonance condition
has the form  \mbox{$k_y=\sigma/2$}. For a wave propagating in a 
boundless medium this condition would lead to an amplitude decay 
\mbox{$\propto \exp[-\mbox{Im} (k_y)y]$}
in the $y$ direction, for light within the frequency range 
$\Delta\omega\approx \sigma c \delta\varepsilon$ (the photonic stop band).
However, the boundary conditions 
enforce a real $k_y$ value (see Fig.~1). Consequently,
instead of causing a finite $\mbox{Im}(k_y)$, the Bragg condition
manifests itself in splitting of the $z$ projection of the wave vector.
This may be seen from the following relation between the components
of the wave vector $\tilde{\mb{k}}$ inside the medium in 
the vicinity of the Bragg resonance:
%
\begin{equation}\label{conservation}
\cos^2\theta_{\scr{B}}\left(\delta\tilde{k}_z\right)^2-
\sin^2\theta_{\scr{B}}\left(\delta\tilde{k}_y\right)^2=
\left(\frac{k\delta\varepsilon}{2}\right)^2,
\end{equation}
where $\delta\tilde{k}_z=\tilde{k}_z-k\cos\theta_{\scr{B}}$ and
$\delta\tilde{k}_y=\tilde{k}_y-k\sin\theta_{\scr{B}}$.
The derivation of Eq.~(\ref{conservation}) is sketched in the Appendix.
With $\tilde{k}_y=k_y=k\sin\theta$ fixed by the boundary conditions,
Eq.~(\ref{conservation}) yields two values of $\tilde{k}_z$, namely 
$\tilde{k}_z=k\cos\theta_{\scr{B}}\pm\Omega$ with
\begin{eqnarray} \label{omegalb} \nonumber
\Omega&=&\frac{1}{2k\cos\theta_{\scr{B}}}\sqrt{
(\sigma k\cos\theta_{\scr{B}}\;\beta)^2+
(k^2\delta\varepsilon)^2} \\
&=&\frac{\tan\theta_{\scr{B}}}{2L_{\scr{B}}}
\sqrt{(2kL_{\scr{B}}\cos\theta_{\scr{B}}\;\beta)^2+1\;},
\end{eqnarray}
where $\beta=\theta-\theta_{\scr{B}}$ is the deviation
from the Bragg angle and 
$L_{\scr{B}}=2\sin^2\theta_{\scr{B}}/(\sigma\delta\varepsilon)$
is the Bragg length. In the region $z>0$, the field components
with $z$ projections $k\cos\theta_{\scr{B}}+\Omega$ and 
$k\cos\theta_{\scr{B}}-\Omega$,
which comprise the waves $\tilde{\mb{k}}$ and 
$\tilde{\mb{k}}-\mb{\sigma}$ shown
in Fig.~1, are coupled to each other.
This leads to the following modification of Eq.~(\ref{psi})
for $\Psi_{\scr{inc}}$:
\begin{eqnarray} \label{psi-rabi} \nonumber
\Psi_{\scr{inc}}(\mb{r})&\approx& \exp\left(
-\frac{z}{l\cos\theta_{\scr{B}}}\right) \\
&\times&\left\{
C_{\Omega,\phi}(z)e^{i\mbscr{k}_{B}\mbscr{r}}
+iS_{\Omega,\phi}(z)e^{i(\mbscr{k}_B-\mbscr{\sigma})\mbscr{r}}
\right\},
\end{eqnarray}
where the functions $C_{\Omega,\phi}(z)$ and
$S_{\Omega,\phi}(z)$ are defined as
\begin{eqnarray} \label{ccoeff}
C_{\Omega,\phi}(z)&=&\cos(\Omega z) - i \cos\phi\sin(\Omega z), \\
\label{scoeff}
S_{\Omega,\phi}(z)&=&\sin\phi\;\sin(\Omega z),
\end{eqnarray}
and $\phi$ is determined by the relation 
\begin{equation}\label{phi}
\sin\phi=\frac{1}{\sqrt{(2kL_{\scr{B}} 
\cos\theta_{\scr{B}}\;\beta)^2+1\;}}.
\end{equation}

The first term of the field amplitude in Eq.~(\ref{psi-rabi}) 
is the wave transmitted
through the interface and traveling along direction close
to the direction $\mb{k}$ of the incident wave,
$(\mb{k}_{\scr{B}})_z=k\cos\theta_{\scr{B}}$, and
$(\mb{k}_{\scr{B}})_{xy}=k_{xy}=k\sin\theta$. The second term is the
diffracted satellite wave. As is seen in Eq.~(\ref{scoeff}), 
the characteristic length scale at which the latter wave
is formed is $\xi=1/\Omega$.
In the limit $L_B\rightarrow\infty$ the incident wave does not change
upon crossing the boundary. Indeed, the
satellite wave in Eq.~(\ref{psi-rabi}) vanishes due to the
$\sin\phi$ prefactor in Eq.~(\ref{scoeff}), whereas $C_{\Omega,\phi}(z)$ 
turns into 
$\exp(-ik\sin\theta_{\scr{B}}\;\beta\; z)$ 
and makes up for the difference between
$(\mb{k}_{\scr{B}})_z$ and $k_z=k\cos\theta$.

The modification due to the Bragg scattering should be incorporated 
into the emerging wave propagator $G$ in a similar way. 
Taking into account that after the last 
\parbox[t]{8.65cm}{
\noindent
\centerline{\psfig{figure=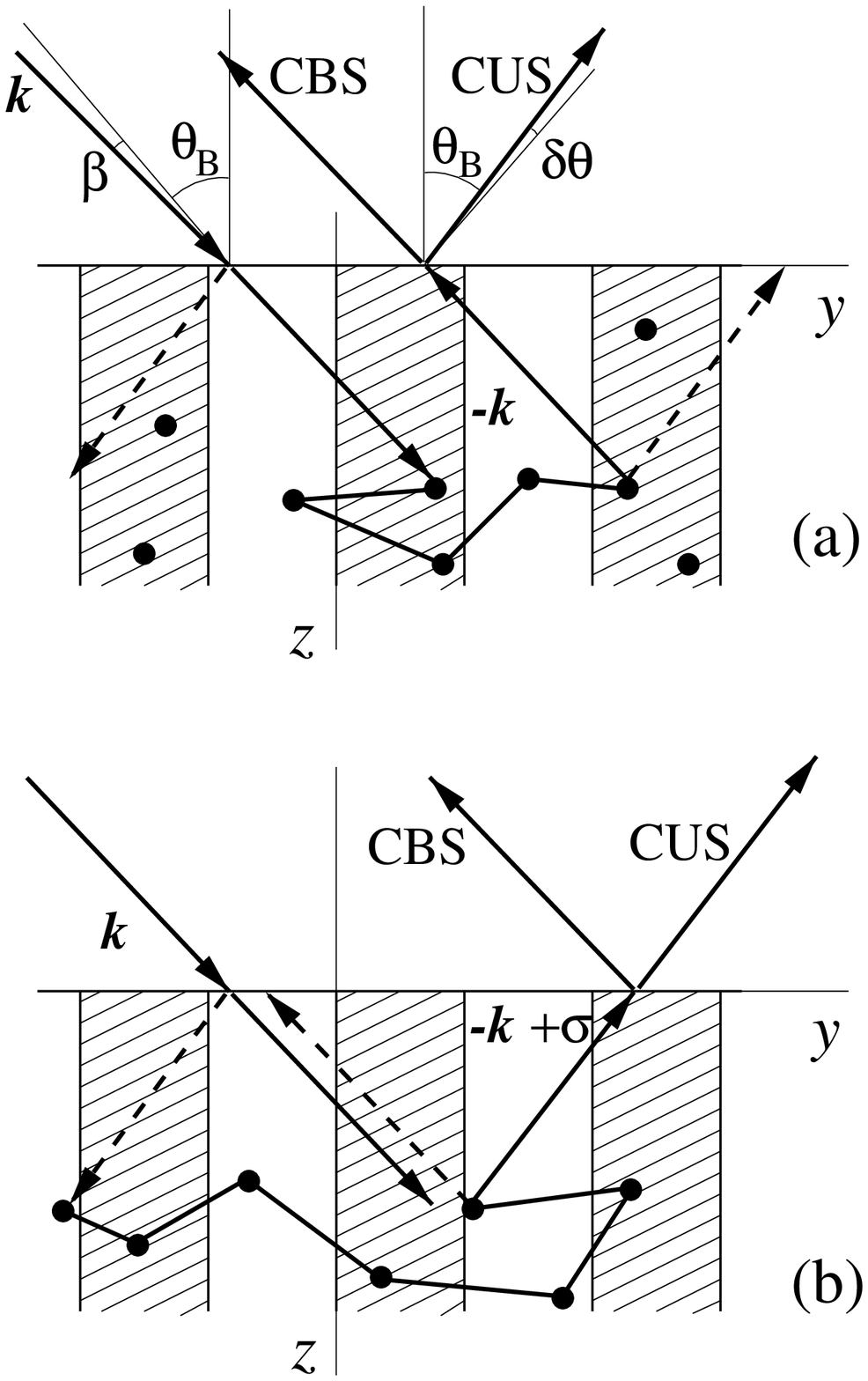,width=8cm}}

\vskip2mm

\noindent
\small {\bf Fig. 2.} Schematic illustration of 
different contributions to the coherent
albedo. CUS peak originates from (a) diffraction of the backscattered 
light and (b) backscattering of the diffracted light.

\vskip3mm
}
scattering event a wave 
$\exp(i\mb{k}'\mb{r})$ that
emerges at angle $\theta'=\theta_{\scr{B}}+\delta\theta$ 
close to the Bragg resonance 
also has a diffracted component inside the medium, Eq.~(\ref{g}) transforms 
into
\begin{eqnarray}\label{g-rabi} \nonumber
& &G(\mb{R},\mb{r})\approx \frac{e^{ikR}}{4\pi R}\exp\left(
-\frac{z}{l\cos\theta}\right) \\
& &\times
\left\{ C_{\Omega',\phi'}(z)e^{-i(\mbscr{k}'_B-\mbscr{\sigma})\mbscr{r}}
+ iS_{\Omega',\phi'}(z)e^{-i\mbscr{k}'_B\mbscr{r}}\right\},
\end{eqnarray}
where $\Omega'$ and $\phi'$ are given by the same 
expressions~(\ref{omegalb}),
(\ref{phi}), where $\beta$ is replaced by 
$\delta \theta$. The components of the wave vector 
$\mb{k}_{\scr{B}}'$ inside the medium are given by 
$(\mb{k}_{\scr{B}}')_z=-k\cos\theta_{\scr{B}}$ and
$(\mb{k}_{\scr{B}}')_{xy}=k'_{xy}=k\sin\theta'$.
Expression~(\ref{g-rabi}) is written specifically for the observation
point in the CUS direction, {\em i.e.} $(\mb{k}_{\scr{B}})_y>0$ (Fig. 1).
Similarly to Eq.~(\ref{psi-rabi}), $G(\mb{R},\mb{r})$
represents the sum of two terms, which we dub here as $C$-term and
$S$-term, respectively. 

\subsubsection{CUS and CBS albedo}

Substitution of Eqs.~(\ref{psi-rabi}) and~(\ref{g-rabi}) 
together with Eqs.~(\ref{ui}) and~(\ref{pdiff}) 
into Eq.~(\ref{albedo}) produces a sum of various terms with 
fast oscillating exponential factors. The terms with factors
$e^{\pm i\mbscr{\sigma}\mbscr{r}_1}$ or 
$e^{\pm i\mbscr{\sigma}\mbscr{r}_1'}$
average out upon integration. Since
Eq.~(\ref{g-rabi}) is valid only 
in the vicinity of the Bragg resonance $\mb{k}'\approx
-\mb{k}+\mb{\sigma}$, then
nonresonant terms should be discarded.
Note however, that, in addition to the oscillating terms,
the product of field amplitudes~(\ref{psi-rabi}) and
propagators~(\ref{g-rabi}) in Eq.~(\ref{albedo}) contains now {\em two}
terms proportional to 
$\exp[i(\mb{k}_{\scr{B}}'+\mb{k}_{\scr{B}}-\mb{\sigma})\mb{\rho}]$,
which {\em do not vanish} if the outgoing light is
parallel to the CUS direction. 
The total CUS albedo is determined by these terms and is thus given by
\end{multicols}
\vskip-7.5mm
\hbox to8.9cm{\hrulefill\vrule height2mm}
\vskip-3mm
\begin{eqnarray} \nonumber
\alpha_{\scr{CUS}}(\mb{k},\mb{k}')&=& \frac{c}{4\pi l^2 S}\;
\int d^2\mb{\rho}dz_1 dz_1'\;  \exp\left[i\mb{q}\mb{\rho}
-b\frac{(z_1+z_1')}{l}\right]\;P(\mb{r}_1,\mb{r}_1') \\ \label{ascint}  
&\times&\left[
C_{\Omega',\phi'}(z_1')C^*_{\Omega,\phi}(z_1')S_{\Omega,\phi}(z_1)
S^*_{\Omega',\phi'}(z_1)+
C_{\Omega,\phi}(z_1)C^*_{\Omega',\phi'}(z_1)S_{\Omega',\phi'}(z_1')
S^*_{\Omega,\phi}(z_1')\right],
\end{eqnarray}
where $\mb{q}=(\mb{k}_{\scr{B}}+\mb{k}_{\scr{B}}'-\mb{\sigma})_{xy}$.

The origin of the two terms in the coherent scattering albedo
is schematically illustrated in Fig.~2. In Fig.~2a the $C$-component of 
the incident light (solid line) {\em first} experiences coherent 
backscattering and {\em then} is diffracted into the $S$ component
(dashed line). 
In Fig.~2b the $C$-component of the incident light is
{\em first} diffracted into the $S$-component, the
backscattering of which provides
the second contribution to the CUS.

The origins of the {\em two} contributions to the CBS can be traced
from Eq.~(\ref{albedo}) in a similar way (see also Fig.~2). 
The only technical difference between the derivations of the CBS and CUS 
is that for the observation point $\mb{R}$ in the CBS direction
({\em i.e.} $\mb{k}'\approx-\mb{k}$), $\mb{\sigma}$ should
be added to the wave vectors in both oscillating
exponents of the Green function in Eq.~(\ref{g-rabi}).

Substitution of the modified propagator 
into Eq.~(\ref{albedo}) and selection of non-vanishing
resonant terms produce the following expression for the CBS
albedo
\begin{eqnarray} \nonumber
\alpha_{\scr{CBS}}(\mb{k},\mb{k}')&=&  \frac{c}{4\pi l^2 S}\;
\int d^2\mb{\rho}dz_1 dz_1'\;  \exp\left[i\mb{q}\mb{\rho}
-b\frac{(z_1+z_1')}{l}\right]\;P(\mb{r}_1,\mb{r}_1') \\ \label{accint}  
&\times& \left[C^*_{\Omega',\phi'}(z_1)C_{\Omega',\phi'}(z_1')
C_{\Omega,\phi}(z_1)
C^*_{\Omega,\phi}(z_1')+
S^*_{\Omega',\phi'}(z_1)S_{\Omega',\phi'}(z_1')S_{\Omega,\phi}(z_1)
S^*_{\Omega,\phi}(z_1')\right],
\end{eqnarray}
\begin{multicols}{2}
\noindent
with $\mb{q}=(\mb{k}_{\scr{B}}+\mb{k}_{\scr{B}}')_{xy}$. 
In the limit
$L_B\rightarrow\infty$, the diffracted waves vanish, 
$S_{\Omega,\phi}\rightarrow 0$, 
so that the only contribution to the
albedo that survives in this limit comes from the
first term of Eq.~(\ref{accint}).

\section{Results and discussion}

\subsection{Analytical results}

\subsubsection{Expressions for CBS and CUS albedo}

The additional oscillating factors $C(z)$ and $S(z)$ in the 
integrands~(\ref{ascint}), (\ref{accint}) compared to Eq.~(\ref{aint})
can be formally absorbed into the decrement $b$ by adding to it 
imaginary parts of the type $i(\Omega\pm\Omega')l$. 
Then all the contributions to
the coherent albedo can be conveniently expressed
with the help of an auxiliary function, $\tilde{f}(\kappa,p)$,
defined as
\begin{eqnarray} \nonumber
\tilde{f}(\kappa,p)&=&f(\kappa,
q,b+ipl)+
f(\kappa,q,
b-ipl)\\ \label{ftildedef}
 &=& \frac{3}{4\pi S D}\;
\frac{1}{1+X^2}
\left[\frac{b-plX}
{b^2+p^2l^2}+\frac{1-e^{-2qz_0}}{ql}
\right]\!,
\end{eqnarray}
where $f(\kappa,q,b)$ is the shape of the CBS cone 
given by Eq.~(\ref{fdef});
the parameters $D$ and $X$ are 
expressed through the arguments of the function
$f$ as follows
\begin{eqnarray}
D=(\kappa^2-p^2)l^2+(b+ql)^2,\\
X=\frac{2pl(b+ql)}{D}.
\end{eqnarray}
In Eq.~(\ref{ftildedef}) the wave vector $q$ is equal to
$q=k\cos\theta_{\scr{B}}|\beta-\delta\theta|$ for
the CBS, and $q=k\cos\theta_{\scr{B}}|\beta+\delta\theta|$ for 
the CUS; in the vicinity of the Bragg resonance we can set 
\mbox{$b=(\cos\theta)^{-1}+(\cos\theta')^{-1}\approx
2(\cos\theta_{\scr{B}})^{-1}$}. Finally, the 
the two contributions to the CBS
albedo in Eq.~(\ref{accint}) take the form
\end{multicols}
\vskip-6.5mm
\hbox to8.9cm{\hrulefill\vrule height2mm}
\vskip-3mm
\begin{eqnarray} \nonumber
\alpha^{(1)}_{\scr{CBS}}&=& \frac{\sin^2\phi\;\sin^2\phi'}{16}\left[
\tilde{f}(0,\Omega+\Omega')+\tilde{f}(0,
\Omega-\Omega')\right]+ \\ \nonumber
& &\frac{(1+\cos^2\phi)(1+\cos^2\phi')}{16}\left[\tilde{f}
(\Omega+\Omega',0)
+\tilde{f}(\Omega-\Omega',0)\right]+ \\ \nonumber
& &\frac{(1+\cos^2\phi)\sin^2\phi'}{8}\tilde{f}(\Omega,\Omega')+
\frac{(1+\cos^2\phi')\sin^2\phi}{8}\tilde{f}(\Omega',\Omega)- \\
\label{acbs1}
& &\frac{\cos\phi\;\cos\phi'}{4}\left[\tilde{f}(\Omega+\Omega',0)-
\tilde{f}(\Omega-\Omega',0)\right], \\ \nonumber
\alpha^{(2)}_{\scr{CBS}}&=&\frac{\sin^2\phi\;\sin^2\phi'}{16}\left[
\tilde{f}(\Omega+\Omega',0)+\tilde{f}(\Omega-\Omega',0)+
\tilde{f}(0,\Omega+\Omega')\right. +\\ \label{acbs2}
& &\left. \tilde{f}(0,\Omega-\Omega')-
\tilde{f}(\Omega',\Omega)-
\tilde{f}(\Omega,\Omega')\right].
\end{eqnarray}
Analogously, the CUS albedo Eq.~(\ref{ascint}) can be expressed
through the function $\tilde{f}$ in the following way 
\begin{eqnarray} \nonumber
\alpha_{\scr{CUS}}&=&\frac{\sin\phi\;\sin\phi'}{8}\left\{
(1+\cos\phi\;\cos\phi')\left[\tilde{f}(0,\Omega-\Omega')+
\tilde{f}(\Omega-\Omega',0)-\tilde{f}(\Omega,\Omega')
-\tilde{f}(\Omega',\Omega)
\right]\right. - \\ \label{acus2}
& &\left. (1-\cos\phi\;\cos\phi')\left[\tilde{f}(0,\Omega+\Omega')+
\tilde{f}(\Omega+\Omega',0)-\tilde{f}(\Omega,\Omega')
-\tilde{f}(\Omega',\Omega)
\right]\right\}, 
\end{eqnarray}
Expressions~(\ref{acbs1})--(\ref{acus2}) are our main results.
Below we analyze two limiting cases of small and large $L_{\scr{B}}$.
%
\vbox{\hspace*{8.5cm}\hbox to8.9cm{\vrule depth2mm\hrulefill}}
\vspace*{-6mm}
\begin{multicols}{2}

\subsubsection{Limiting cases}

\vskip-1mm

Let us trace how the conventional CBS cone is recovered in
the limit $L_B\rightarrow\infty$. In this limit we have
$\phi,\phi'\rightarrow 0$, so that
$\alpha_{\scr{CUS}}$ and 
$\alpha^{(2)}_{\scr{CBS}}$ containing $\sin\phi$ and/or $\sin\phi'$ 
as prefactors, vanish. Substituting $\phi=\phi'=0$ into 
Eq.~(\ref{acbs1}) we obtain
$\alpha_{\scr{CBS}}^{(1)}=f(\Omega-\Omega',q,b)$. 
Taking the limit $L_{\scr{B}}\rightarrow\infty$ in
Eq.~(\ref{omegalb}), we get 
$\left(\Omega-\Omega'\right)\rightarrow 
k\sin\theta_{\scr{B}}\;(\beta-\delta\theta)$. Correspondingly,
$f(\Omega-\Omega',q,b)$ reduces to Eq.~(\ref{fdef}). 

Consider now the opposite limit\cite{Gorodnichev}, 
$L_{\scr{B}}/l\ll 1$.
In this limit $\phi,\phi'\rightarrow\pi/2$. It follows from
Eq.~(\ref{omegalb}), that with decreasing $L_{\scr{B}}$
both $\Omega$ and $\Omega'$ diverge, 
while $\left(\Omega-\Omega'\right)\rightarrow 0$. As a result,
as can be also seen from Eq.~(\ref{ftildedef}),
all of the terms in Eqs.~(\ref{acbs1})--(\ref{acus2}) that contain
$\Omega$, $\Omega'$ or $\Omega+\Omega'$ as at least one 
of the arguments, vanish. Then
it is straightforward to check that 
both $\alpha_{\scr{CBS}}$ and
$\alpha_{\scr{CUS}}$ take the form 
$\tilde{f}(0,0)/4=f(0,q,b)/2$. We thus recover the result of 
Ref.\onlinecite{Gorodnichev} that in the limit of strong modulation, 
CBS and CUS cones are the mirror images of each other. Concerning
the shape of the cones, 
it is given by the conventional expression~(\ref{fdef}) for the
coherent albedo. Concerning the peak heights, they are 
two times less
than the height of the CBS peak in the absence of modulation.

\subsection{Shapes of the CBS and CUS peaks}

Expressions~(\ref{acbs1})--(\ref{acus2}) for the CBS
and CUS albedo are valid
only in the vicinity of the Bragg resonance 
\parbox[t]{8.65cm}{
\noindent
\centerline{\psfig{figure=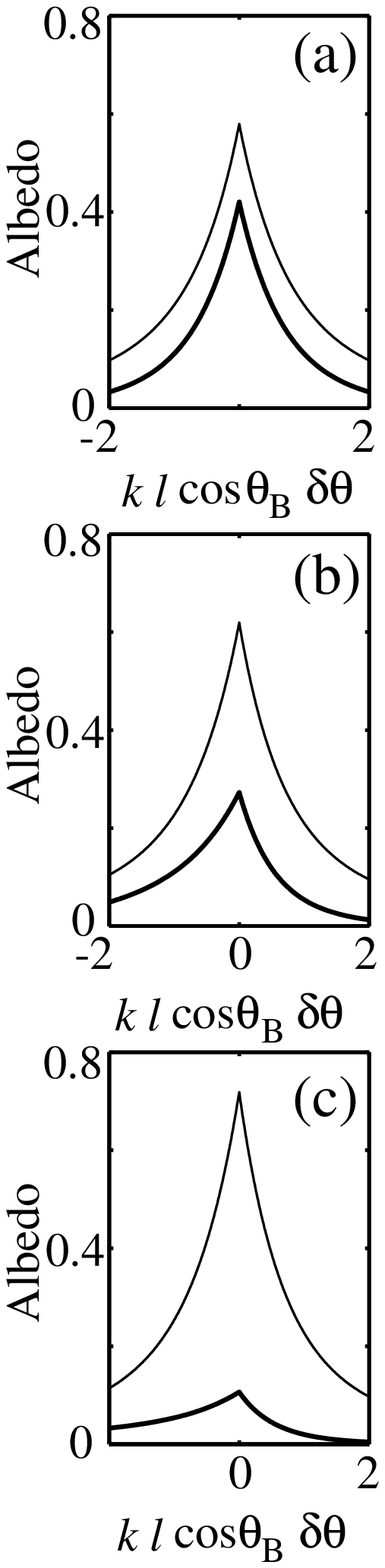,height=15cm}}

\vskip2mm

\noindent
\small {\bf Fig. 3.} CBS (thin line) and CUS (bold line) peaks 
at  $L_{\scr{B}}=0.3l$ normalized
to the CBS peak height at $L_{\scr{B}}/l\rightarrow\infty$ are plotted
for detuning (a) $\beta=0$;
(b) $|\beta|=(kl\cos\theta_{\scr{B}})^{-1}$;
(c) $|\beta|=2(kl\cos\theta_{\scr{B}})^{-1}$. 
The peak maxima occuring at $\delta\theta=\beta$ 
($\delta\theta=-\beta$) for the CBS (CUS) are shifted
to $\delta\theta=0$ for convenience.

\vskip3mm
}
$\theta,\theta'\approx\theta_B$. Note, however, that only in 
this region CUS has an appreciable amplitude.
To illustrate this we plot in Fig.~3 the CBS and CUS cones
for different detunings, $\beta=\theta-\theta_{\scr{B}}$, of the
incident beam from the Bragg 
angle for $L_{\scr{B}}/l=0.3$. It is seen that CUS
practically dies out at 
$\beta\approx(2kL_{\scr{B}}\cos\theta_{\scr{B}})^{-1}\ll 1$.
It is also seen that as the amplitude of the CUS peak falls off,
the peak also becomes asymmetric.
The behavior of the CUS and CBS peak heights with the detuning $\beta$
is summarized in Fig.~4.

Consider now the case of exact resonance, $\beta=0$.
As was discussed above, the relation between the CBS and
CUS peaks is governed by the dimensionless parameter $L_{\scr{B}}/l$.
In Fig.~5 we show the heights 
of both peaks
as functions of $L_{\scr{B}}/l$. It is seen that the amplitude
of the CBS
\parbox[t]{8.65cm}{
\noindent
\centerline{\psfig{figure=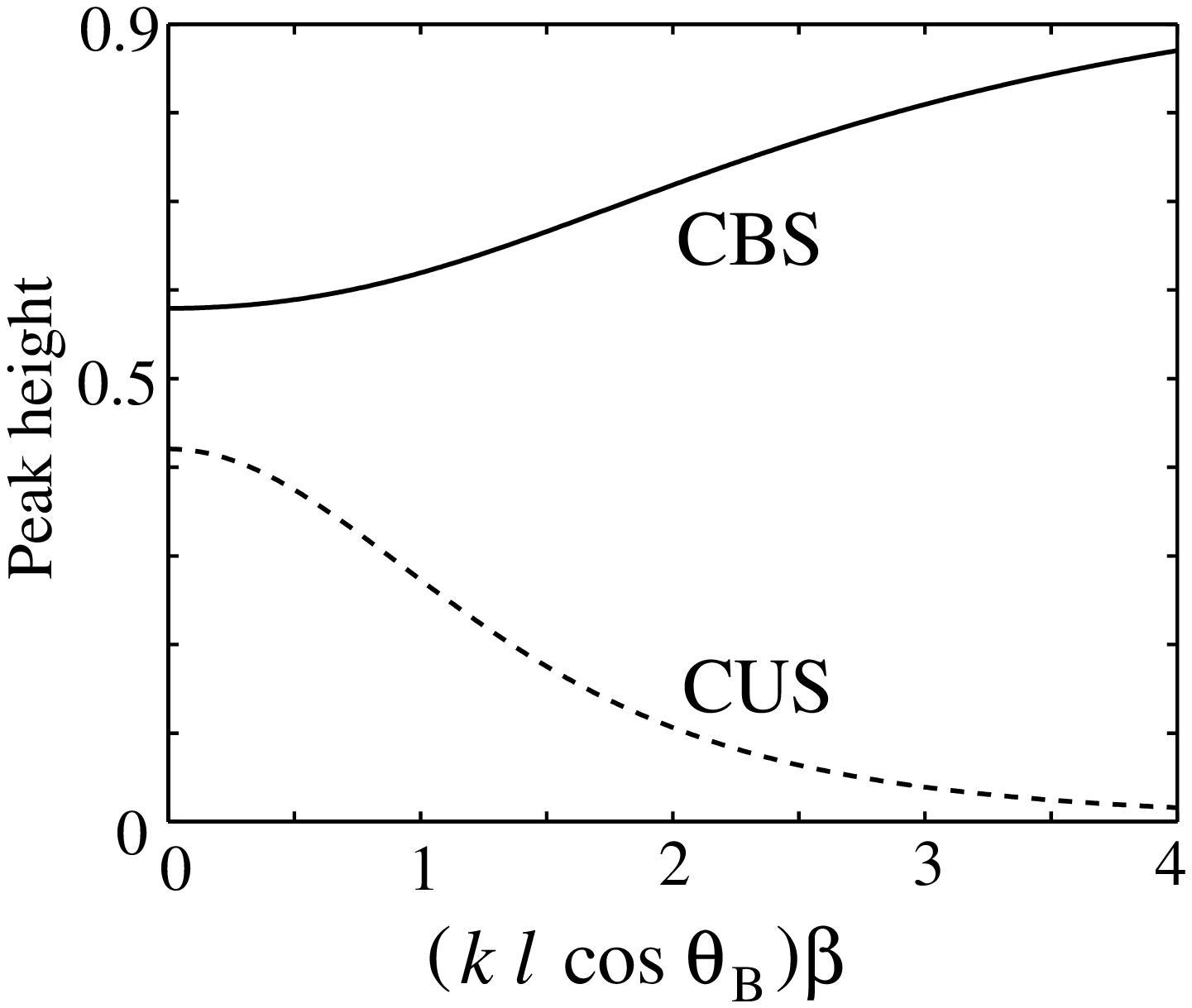,width=8.65cm}}

\vskip2mm

\noindent
\small {\bf Fig. 4.}  CBS and CUS peak heights normalized to the 
CBS peak height
at $L_{\scr{B}}/l\rightarrow\infty$ plotted vs. the 
detuning $(kl\cos\theta_{\scr{B}})\beta$ at $L_{\scr{B}}=0.3l$.

\vskip3mm
}
 peak saturates already at 
$L_{\scr{B}}/l\gtrsim 2$; at the same time,
the CUS peak diminishes by an order of magnitude.

In Fig.~6 we illustrate the evolution of the
CUS cone at $\beta=0$ with increasing
$L_{\scr{B}}/l$. For reference the CBS cone for 
$L_{\scr{B}}/l=0.3$ is also plotted in Fig.~6 (solid line). We note
that, while the CUS peak keeps narrowing for 
$L_{\scr{B}}\lesssim l$, the shape of the CBS cone
remains practically unchanged 
with $L_{\scr{B}}/l$.

\section{Conclusions}


In this paper we have studied coherent light scattering from 
a disordered photonic crystal with incomplete band gaps. We have
demonstrated that the crystal dielectric function periodicity gives rise
to additional coherent albedo peaks in non-backscattering
directions. These peaks emerge as the angle of incidence,
$\theta$, approaches the Bragg resonance, $\theta\approx\theta_{\scr{B}}$.
For simplicity, 
the consideration was restricted to the case where the modulation wave
vector, $\mb{\sigma}$, is parallel to 
the crystal boundary. In this case, and under the Bragg resonance condition
$2k\sin\theta_{\scr{B}}=\sigma$, the direction of the additional
(CUS) peak coincides with the reflection direction, since
the CUS wave vector is equal to 
$\mb{k}'=-\mb{k}+\mb{\sigma}$. The component of $\mb{k}'$ parallel to
the boundary is equal to $k'_y=-k\sin\theta+\sigma$. Then at 
$\theta=\theta_{\scr{B}}$ we have $k_y'=k_y=\sigma/2$. It is important
to note that this relation is specific for $\mb{\sigma}$ parallel
to the boundary (and for the wave vector of incident light lying
in $yz$ plane). 

Consider now the case where $\mb{\sigma}$ and the
incident light wave vector still lie within $yz$ plane, but $\mb{\sigma}$
is no longer parallel to the boundary, but rather makes an angle
$\gamma$ with the $y$ axis. In this case the Bragg condition
takes the form $2k\sin\tilde{\theta}_{\scr{B}}=\sigma$, where
$\tilde{\theta}_{\scr{B}}=\sin^{-1}\left(\frac{\sigma}{2k}\right)-
\gamma$. 
\parbox[t]{8.65cm}{
\noindent
\centerline{\psfig{figure=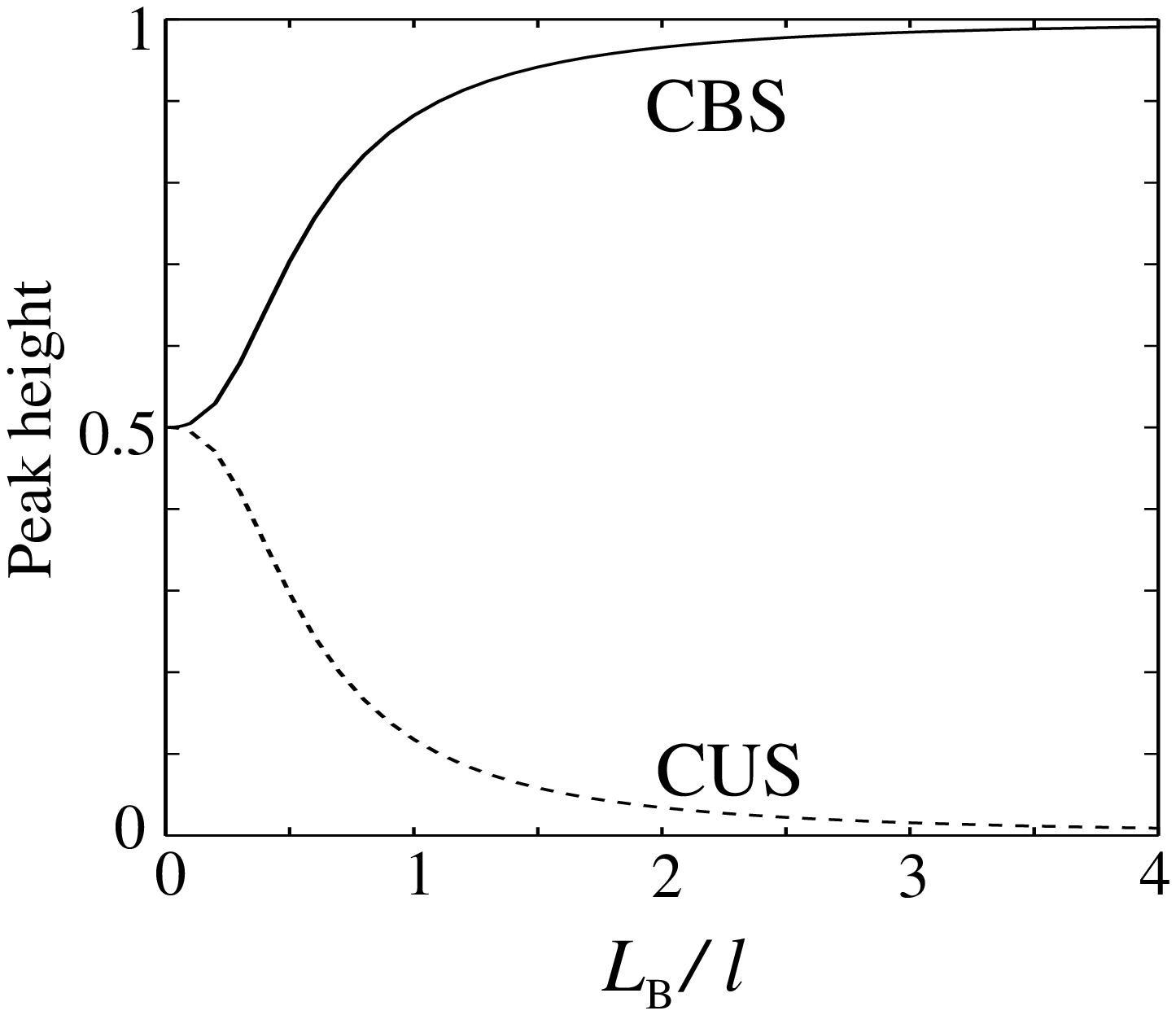,width=8.65cm}}

\vskip2mm

\noindent
\small {\bf Fig. 5.}  CBS and CUS peak heights normalized to the 
CBS peak height
at $L_{\scr{B}}/l\rightarrow\infty$ plotted vs. $L_{\scr{B}}/l$.

\vskip3mm
}
It is easy to see that together with the CUS condition
$\mb{k}'=-\mb{k}+\mb{\sigma}$ this determines the CUS direction
$\theta_{\scr{CUS}}=\tilde{\theta}_{\scr{B}}+2\gamma$, whereas
the reflection angle is $\theta_r=\tilde{\theta}_{\scr{B}}$.
This latter case can be exploited in experiments.

Generalization of our theory to the case of finite $\gamma$
amounts to the replacement of the functions $C$ and $S$ in 
Eqs.~(\ref{ascint}), (\ref{accint}) for the CUS and CBS by
\begin{eqnarray} \label{ccoeffgamma}
C_{\tilde{\Omega},\tilde{\phi}}&=&\cos\tilde{\Omega}z-i\sin\tilde{\phi}
\sin\tilde{\Omega}z, \\ \label{scoeffgamma}
S_{\tilde{\Omega},\tilde{\phi}}&=&
\left[\frac{\cos\tilde{\theta}_{\scr{B}}}{\cos(\tilde{\theta}_{\scr{B}}+
2\gamma)}\right]^{1/2}\!\!
\sin\tilde{\phi}\sin\tilde{\Omega}z,
\end{eqnarray}
where 
\begin{equation} \label{Omegagamma}
2\tilde{\Omega}=\frac{\sin\theta_{\scr{B}}}{\tilde{L}_{\scr{B}}
\cos(\tilde{\theta}_{\scr{B}}+2\gamma)}
\sqrt{(2k\tilde{L}_{\scr{B}}\cos(\tilde{\theta}_{\scr{B}}+\gamma)\;
\beta)^2+1\;}\!,
\end{equation}
is the modified splitting between the two solutions 
of Eq.~(\ref{conservation}) for $k_z$. 
The modified Bragg length in Eq.~(\ref{Omegagamma}) is defined as
$\tilde{L}_{\scr{B}}=L_{\scr{B}}\sqrt{\cos\tilde{\theta}_{\scr{B}}/
\cos(\tilde{\theta}_{\scr{B}}+2\gamma)}$. The parameter
$\tilde{\phi}$ in Eq.~(\ref{Omegagamma}) is still given
by Eq.~(\ref{phi}) with $L_{\scr{B}}\rightarrow\tilde{L}_{\scr{B}}$.

To derive Eqs.~(\ref{ccoeffgamma})--(\ref{Omegagamma}) it is
convenient to perform a rotation of the coordinate system 
by the angle $\gamma$. Then the additional factors
in~(\ref{scoeffgamma}), (\ref{Omegagamma}) as compared to
$\gamma=0$ emerge due to the modification of
the boundary conditions. Other steps of the derivation
remain unchanged.

It is seen from Eqs.~(\ref{scoeffgamma}), (\ref{Omegagamma}) that
the prefactors diverge at
$2\gamma=\frac{\pi}{2}-\tilde{\theta}_{\scr{B}}$.
This divergence corresponds to the physical situation where
the diffracted component of the incident light is aligned
with the boundary. With regard to the coherent scattering, the 
condition $2\gamma=\frac{\pi}{2}-\tilde{\theta}_{\scr{B}}$
manifests a crossover to a new regime.
\parbox[t]{8.65cm}{
\noindent
\centerline{\psfig{figure=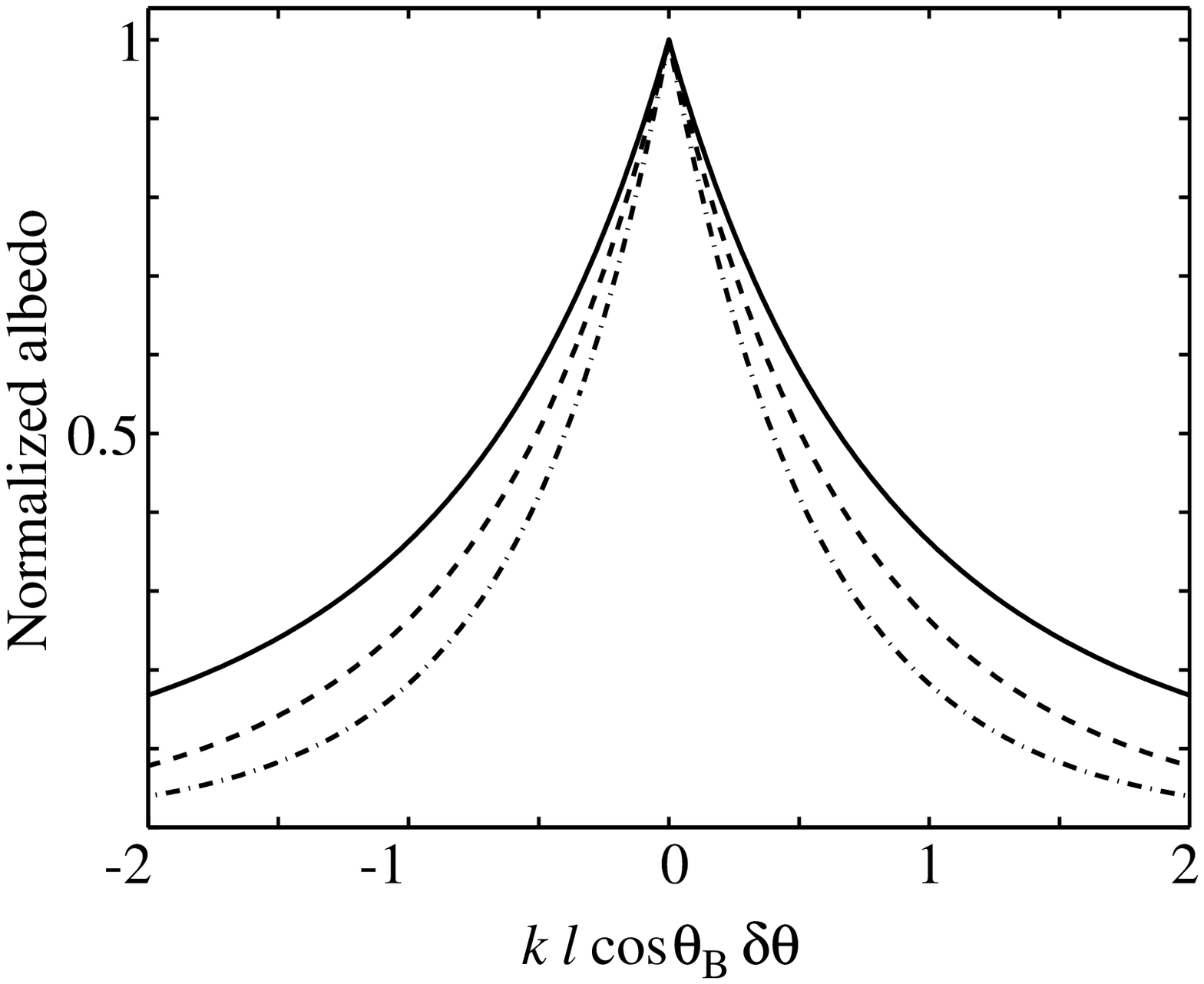,width=8.65cm}}

\vskip3mm

\noindent
\small {\bf Fig. 6.} Normalized CUS albedo at exact resonance, 
$\beta=0$, shown for  $L_{\scr{B}}=0.3l$ (dashed line) 
and $L_{\scr{B}}=l$ (dot-dashed line).
Solid line: normalized CBS albedo at 
$\beta=0$ and $L_{\scr{B}}=0.3l$.
\vskip2mm
}
In this regime the diffracted wave does not ``fit'' into the
medium, so that the CUS peak is absent. Formally,
for  $2\gamma>\frac{\pi}{2}-\tilde{\theta}_{\scr{B}}$
Eq.~(\ref{conservation}) does not have real solutions
for $k_z$ for small $\beta$, 
which corresponds to
opening of the photonic band gap in $z$ direction.  
As a result, the CBS peak exhibits an anomaly in the vicinity
of the Bragg condition\cite{Raikh}. 

We did not address in the present work the modifications of the theory
caused by a difference of refraction indices on two sides of the
interface. The impact of the light refraction on the CBS was
intensively studied (see Ref.~\onlinecite{review} and references
therein). A non-trivial consequence of the refraction index 
mismatch is that in course of diffusion the light wave can ``strike'' the
boundary at an angle exceeding the angle of total internal reflection. 
Basically, the CUS can be viewed as a mirror image of the CBS, hence the
effect of the index mismatch on CUS is similar to that on CBS.

{\bf Acknowledgments}. This work was supported in part by 
the
Petroleum Research Fund under grant ACS-PRF \#34302-AC6, and the Army
Research Office grant DAAD 19-0010406

\appendix
\section*{}

In the vicinity of the Bragg resonance, $k_y\sim\sigma/2$, we
employ the coupled wave approach, {\em i.e.}
we search for the field inside
the medium with dielectric function given by Eq.~(\ref{epsilon}) in the
form of a sum of two waves with wave vectors $\mb{k}$
and $\mb{k}-\mb{\sigma}$ and amplitudes 
$A_{\mb{k}}$ and $A_{\mb{k}-\mb{\sigma}}$, respectively. 
Substitution of this form into the wave
equation results in a following system of coupled equations
for the wave amplitudes 
\end{multicols}
\begin{equation}\label{secular}
\begin{array}{l}
\left[(k\cos\theta_{\scr{B}})^2-k_z^2-2k\sin\theta_{\scr{B}}\;
(k_y-k\sin\theta_{\scr{B}})\right] A_{\mb{k}} +k^2\delta\varepsilon \;
A_{\mb{k}-\mb{\sigma}} = 0\\
k^2\delta\varepsilon\;A_{\mb{k}}+\left[(k\cos\theta_{\scr{B}})^2-k_z^2+
2k\sin\theta_{\scr{B}}\;(k_y-k\sin\theta_{\scr{B}})\right]
A_{\mb{k}-\mb{\sigma}}=0,
\end{array}
\end{equation}
\vbox{\hspace*{8.7cm}\hbox to8.85cm{\vrule depth2mm\hrulefill}}
\vspace*{-6mm}
\begin{multicols}{2}
where we used the definition of the Bragg angle
$\sin\theta_{\scr{B}}=\sigma/(2k)$.
The system~(\ref{secular})  together
with the assumption  $\cos\theta\approx\cos\theta_{\scr{B}}$
lead to Eq.~(\ref{conservation}).

It is convenient
to express the modulation strength $\delta\varepsilon$
in terms of the Bragg decay length, $L_{\scr{B}}$, given by
\begin{equation} \label{lbdef}
L_B=\frac{2\sin^2\theta_B}{\sigma\delta\varepsilon}.
\end{equation}
It is easy to see from Eq.~(\ref{lbdef}) 
that the meaning of $L_B$ is the decay length 
for light of frequency $\omega=kc$ in the middle of the Bragg gap,
when the medium boundary
is perpendicular to the modulation direction $y$.

If the condition given in Eq.~(\ref{conservation}) is satisfied, then there
exists a nontrivial
solution of the system~(\ref{secular}),
\begin{equation}
\lambda_{1,2}=\frac{A_{\mb{k}-\mb{\sigma}}}{A_{\mb{k}}}=
\frac{\cos\phi\pm1}{\sin\phi},
\end{equation}
with $\phi$ defined by Eq.~(\ref{phi}). The two signs in the numerator
correspond to the two solutions of Eq.~(\ref{conservation}) for $k_z$.
Upon matching the solution for the field amplitude
with the incident wave, the resulting field can be cast in the 
form of Eq.~(\ref{psi-rabi}).

\end{multicols}

\end{document}